\documentclass[fleqn,10pt]{wlscirep}
\usepackage[utf8]{inputenc}
\usepackage[T1]{fontenc}
\usepackage{graphicx, wrapfig}

\graphicspath{ {./images/} }

\title{The Participation Game: \newline A Post-Turing Frontier for Generative AI, \newline with Implications for Theory and Society}

\author[1,*]{Mark Thomas Kennedy}
\author[2]{Nelson Phillips}

\affil[1]{Imperial College London, Data Science Institute, London, SW7 2AZ, United Kingdom}
\affil[2]{University of California Santa Barbara, Technology Management, Santa Barbara, 93106, United States}

\affil[*]{mark.kennedy@imperial.ac.uk}

\affil[+]{these authors contributed equally to this work}

\keywords{AI, ontology, social construction}

\begin{abstract}
Inspired by Turing’s famous "imitation game” and recent advances in generative pre-trained transformers \cite{openai2023gpt4}, we pose “the participation game” to point to a new frontier in AI evolution where machines will join with humans as participants in social construction processes. The participation game is a creative, playful, competition that calls for applying, bending, and stretching the categories humans use to make sense of and order the world. After defining the game and giving reasons for moving beyond imitation as a test of AI, we highlight parallels between the participation game and processes of social construction---a hallmark of human intelligence. We then discuss how having artificial participants in reality-making processes holds implications for theory and society that demand new approaches to AI governance. 
\end{abstract}

\begin{document}

\flushbottom
\maketitle
\thispagestyle{empty}

\section*{Introduction}

When Turing \cite{Turing} asked, “Can machines think?”, he proposed “The Imitation Game”, a test that used general conversational abilities as a proxy for thinking. As generative AI systems deliver increasingly capable conversational abilities, however, scholars remain  that Turing's test was too easy a standard for saying that a machine is thinking. 

In response to new technologies\cite{Attention}\textsuperscript{,}\cite{BERT}\textsuperscript{,}\cite{GPT3} and Searle's\cite{Searle_MBP} critiques of Turing, we ask a different question, “Can machines join with humans in co-creating the world?” Inspired by Turing's imitation game, we propose the participation game, a game in which computers join with people in the sorts of conversations that shape the categories people use to enact and make sense of the world. This challenge pushes computer scientists to engage two fundamentals of human intelligence: all knowledge about reality is socially constructed\cite{BergerandLuckmann}, and the social realities of human cultures are human-made \cite{Searle_CoSR}. Besdies being a hallmark of human intelligence, the capacity to participate in social construction processes is the very thing that enables humans to enact many types of order—including organizations—that enact and structure the social realities of societies\cite{WeickSensemaking}. Even if computers can exert influence in these processes without developing human-like thinking and understanding, having efficacious artificial participants (APs) in reality-making processes will demand both new approaches to AI governance and new theorizing about influence and influencers in public discourse. The participation game offers a way to assess AP capacity and explore the theoretical and practical implications of having APs in social construction. 

\section*{The Participation Game}

The ‘participation game’ builds on a parlour game called \emph{Categories}\cite{Parlett} and a variant called \emph{Scattergories}. In \emph{Categories}, four to six participants compete against a clock and each other to generate a unique word for each of a dozen or so categories, where each word must start with a letter drawn at random -- often by rolling a many-sided die. For example, if the drawn letter is ‘f’ and categories include, for example, foods, places, first names, films, fowl, and colors; one could say fruit, France, Frank, Fargo, flamingos, and fuchsia. When time is up, participants share their lists to seek approval that their words match the categories; when words are debated, approval is decided by majority vote. Participants score 2 points for unique approved words, 1 point for approved words others also wrote, and 0 for words rejected in voting. Play proceeds for a fixed period (e.g., half an hour) or until any player reaches a victory threshold (e.g., 21 points). At the end of the game, the highest score wins.

In our view, the game’s success has much to do with incentives for words that show creativity not only in stretching or reinterpreting categories, but also in the lively discussions and arguments that follow. When games are played at gatherings where not everyone plays, onlookers often heckle participants and disagree or side with whomever they find convincing. In any case, the approval process features explanation, argumentation, and negotiation about concepts and ontologies.

Building on \emph{Categories}, we can explain the participation game succinctly, as follows: play \emph{Categories} with four to six participants, at least one of whom is an artificial participant (AP), where all participants must be truthfully identified as human or artificial from the start. As with \emph{Categories}, play proceeds for the prearranged period or until any player reaches the agreed point threshold for victory, at which point the highest score wins.

For an AP to win, it must be like any successful player: creative in the words it comes up with for each category, persuasive in its arguments for why they should count. Also, the AP will have to be convincing in its critiques of other players’ words and arguments for their words. Like other players, the AP can win by simply getting the highest score. We like this feature of the game because it reflects our view that APs can contribute to social construction processes without being mistaken for humans, either by subterfuge or confusion. In contrast to Turing's imitation game, the goal is not for computers to pass as human, but to be influential with humans despite being known as computers. 

In Turing’s game, communication takes place via typed text akin to chat interfaces now ubiquitous. That familiar interface is a good baseline, but vocal inflections, facial expressions, and physical gestures are all vital dimensions of human connection and persuasion. We suggest the participation game should evolve through levels ranging from typed chat (level 1) to audio chat (level 2) to video chat (level 3) to virtual and augment reality gatherings (level 4) to gatherings with humanoid robots (level 5).

\section*{Why for Raise the Bar on AI?}

We believe there are three reasons for raising the bar on AI: (1) scholars argue that the Turing Test is not hard enough to confirm that machines can think, (2) chatbot encounters lead many to see the Turing Test as already having been passed, and (3) the spread of human-AI interactions that mimic human teamwork raises questions about how to manage human-AI collaborations.

\subsection*{Not Hard Enough}

Influential academic arguments suggest the Turing Test is too low a standard for declaring machines intelligent. French\cite{French90} and Searle\cite{Searle_MBP} argue, for example, it can be passed without thinking or understanding. In our view, these arguments are persuasive. 

\subsection*{Already Passed}
In the years from the first Loebner Prize competition of the early 1990s\cite{Epstein} to the last in 2020, winning chatbots became increasingly impressive. Dreyfus\cite{Dreyfus} argues the Turing Test was needed to propel this kind of progress. With this progress, the public increasingly accept a shift to chatbot-delivered service transactions.

In the last decade, advances in search and machine learning are enabling new levels of human-computer interaction via natural language. Increasingly, researchers and system builders are embracing “the bitter lesson”\cite{Sutton} that the relentless scale-up of search and learning methods renders explicit knowledge modelling all but obsolete. Conceptually, these advances reflect the Zellig Harris argument\cite{Harris} that language has a “distributional structure” because words “do not occur arbitrarily relative to each other” but instead in “certain positions relative to certain other elements”. Harris’ distributional hypothesis means the neighborhoods of words reveal their meanings, and that makes it possible to use maps of neighborhoods to fill in blanks—empty spots on the street, so to speak. Thus, the distributional structure of language\cite{sahlgren2008distributional} explains how machine learning could create so-called large language models (LLMs) in which the meanings of words are relatively small, relatively dense vectors\cite{ULMFiT,BERT}. As LLMs become more capable, they are evolving from guessing missing words to sensibly filling in larger gaps to generating sentences and paragraphs, writing essays, and interactive conversation, especially in a question-and-answer format. These advances are enabling new practical research in many fields and schemes for assessing their capabilities that echo principles of Turing’s imitation game, albeit without deception.

As LLMs continue to evolve, OpenAI’s Generative Pre-Trained Transformers illustrate the practical implications of the distributional hypothesis (see xGPT\cite{GPT}\textsuperscript{,} \cite{GPT2}\textsuperscript{,} \cite{GPT3}). As LLMs are augmented by adding the ability to recognize still images and interpret moving images\cite{GATO}, some are hailing the systems’ apparent reasoning abilities as precursors of artificial general intelligence (AGI)—the vaunted but elusive goal of AI that approximates human abilities to converse and answer questions on a wide range of topics. Since the mid-twentieth century, AGI predictions have flip-flopped between just-around-the-corner and decades-away. Although we believe AGI is still some ways away, we regard recent LLM advances as something new, and they are attracting journalistic coverage\cite{NYTonLLMs}. Returning to the neighborhood analogy for the distributional hypothesis, LLMs work because turn-taking in conversation is like going from a few model homes to the neighborhood built around them. 

\subsection*{Need to Manage Human-AI Teamwork}
Finally, raising the bar for evaluating AI reflects the fact that humans are increasingly following and relying on AIs. Take the world of massive multiplayer online (MMO) games, for example: it has become a site for innovation in and learning about whether and how humans and AIs can work—and play—together \cite{SourmelisetalMMORPGs}. Besides being great environments for trying out and proving advances in reinforcement learning \cite{Justesenetal19,SilveretalGo}, Suarez et al.\cite{SuarezetalNeuralMMO} argue that MMOs come closer than other type of games or simulated environments to modelling real-world learning contexts. In analyzing trust between human players and AI characters in a large MMO, for example, Ahmad et al.\cite{Ahmadetal_MMOs}  show that human-to-AI trust networks develop and echo the effects of homophily prominent in human-to-human trust networks, but with differences arising from relationships and identities players maintain outside the game. In an MTURK study of human players’ perceptions of human-AI teaming in MMOs, Zhang et al.\cite{ZhangetalAITeammates} respondents expressed a preference for playing with an AI teammate over an unknown human player if the AI teammate could help them win. Interestingly, players express similar expectations of human and AI teammates.

More generally, studies of human-AI interactions suggest people feel better about themselves when non-human assistance (from AI or robots) is framed as autonomous and as having emotions, especially if people can incorporate the benefits of that assistance into their self-concepts\cite{ONeill}. To us, this suggests people can feel good about themselves when they are part of an effective team, even if the team includes AIs in human-like roles. By this logic, humans may well be able to accept AIs as APs in social construction processes like those modeled in our participation game.

\section*{If Raising the Bar, Why the Participation Game?}

The participation game raises the bar for AI to include social interactions that require not only creative reinterpretation of ontologies, but also potential extensions, too. Rather than arguing that machines capable of winning a participation game are thinking, understanding, and feeling as humans do, we argue that having machines participate in social construction processes will present profound challenges to social theory and society. To make that case, we define social construction processes and motivate their to human activity; we note the centrality language, ontologies, categories. and categorization in social construction processes; and we highlight advances that suggest APs may be possible sooner rather than later.

\subsection*{Importance of Social Construction Processes}
Social construction processes underlie all shared representations about reality. In their landmark book \emph{The Social Construction of Reality}, Berger and Luckmann\cite{BergerandLuckmann} explain the “processes by which any body of ‘knowledge’ comes to be socially established as ‘reality’”\cite{BergerandLuckmann}. They argue that language records ‘habitualization’ in the recognized patterns, or ‘typifications’, institutionalized in particular cultural contexts. In this tradition \cite{PowellandDiMaggio}, institutionalization follows the meaning of the Latin verb for “to establish”. When a shared body of knowledge is constructed and transmitted across generations, children receive it “as a given reality confronting the individual in a manner analogous to the reality of the natural world”, and “language appears to the child as inherent in the nature of things”\cite{BergerandLuckmann}. Thus, social construction processes enable people to reason together about realities, and they also shape what can be learned.

As Kuhn\cite{Kuhn} argues, the history of science shows the power and persistence of shared understandings of the world. In the natural sciences, objects of study do not depend on human knowledge or assent, and non-social realities push back against errant theories. Still, Kuhn argues even scientists do not “reject paradigms because confronted with anomalies or counterinstances”\cite{Kuhn}. In other words, the power of socially constructed frameworks is such that prediction failures do not guarantee knowledge updating. This dynamic leads Berger and Luckmann to speak of the social construction of reality, not just knowledge about it.

Nonetheless, Searle\cite{Searle_CoSR} makes a powerful argument about the construction of what he calls social realities---facts of life that \emph{do} depend on human knowledge and assent. Whereas Berger and Luckmann defined reality as that which exists without depending on human knowledge or intent, Searle distinguishes social realities as a distinct class of realities instituted in language and through the transmission of shared knowledge about how the world works. Unlike knowledge about phenomena like gravity, where dead-wrong theories are contradicted with force that can be brutish\cite{Searle_CoSR}, social realities may be reproduced for long periods of time even when they are culturally and historically contingent states of affairs that many regard as sub-optimal.

In our proposed participation game, winning takes more than quickness in coming up with words for categories that start with the same letter: it takes creativity and persuasion in the free-wheeling arguments about more creative interpretations of categories. This creative, persuasive, interpretive use of patterns and symbols is at the heart of social construction processes, and it is essential to what it means to think as humans do about the world not only as it is, but also as it could be.

\subsection*{Centrality of Language, Ontologies, and Categories}
Language is central to human social construction processes: it is both the record of our shared representations of reality and the toolkit for working them out. In Rorty’s\cite{Rorty} philosophy of language, language is used not only in a translation process that spreads shared beliefs about reality, but also in an evaluation process that settles, among other things, collective agreements about the ontological status of new things. For Berger and Luckmann\cite{BergerandLuckmann}, language is central to ‘objectivation’, the process whereby humans assign signs, or names, to patterns people of particular communities find it useful to recognize and reason with as elements of their realities, both social and non-social. In Searle’s analysis of social ontologies, language records the institutionalization of “status functions” that, when accepted and used to reason about the world, become “vehicles of power” to the extent that they establish “obligations, rights, responsibilities, duties, entitlements, authorizations, permissions, requirements, etc.”. Searle speaks of the “deontic powers” of language to explain its essential role in institutionalizing such social realities. To construct a deontological theory of the right and wrong of the way things work in any world, one needs language to set it up. As Searle\cite{Searle_SOSBP} puts it, “no language, no deontology.” 

Thus, languages reflect shared views of the world and institutionalize social ontologies, the collections of  categories humans create to make sense of patterns and “anomalies”\cite{WeickSensemaking} deemed significant enough deserve attention, explanation, and names\cite{KennedyandFiss}. In a study of the kinds of organizations people can recognize, Ruef\cite{Ruef} defines ontologies as “systems of categories, meanings, and identities within which actors and actions are situated.” Following Berger and Luckmann and Searle, we say every category, or entry, in an ontology is socially constructed even if it is for something, like gravity, that does not owe its existence to human intention or recognition. Whereas Searle uses “social ontologies” to speak of things that depend on human recognition and intention, social scientists use the phrase to flag that ontologies are inherently social even for phenomena that do not depend on human knowledge or assent. To us, both are right.

As humans use language to create and share ontologies, categories are “continuously remade, refreshed, and/or maintained, with a lot of skilled work by multiple actors with various interests”—some of whom, in the future, may be artificial intelligences. As we will argue, teams, organizations, and societies face profound changes as AI evolves to gain the culturally oriented intelligence required to join humans in social construction processes.

\subsection*{Prospects for Artificial Participants}

Only a short while ago, the suggestion that an AI might join humans in shaping their realities would have seemed both unlikely and undesirable, but this is changing rapidly. As described earlier, techniques for building large-language models have enabled radical advances in the conversational abilities of AI systems. In the last few years, referring to these systems as “an AI” or “AIs” has become increasingly common, and some researchers have begun to feel that systems seem sentient. For a longer time, humans have been using machines and tools like twitter bots in culture wars where rivals compete to establish or dismiss new social realities. In recent years, climate science and public health have been major fronts in ongoing culture wars, and AI is increasingly being weaponized for use in these skirmishes.

\section*{Implications for Theory and Society}

Whilst the participation game pushes AI to join activities arguably at the upper end of the range of human thought, we do not argue that winning will qualify AIs as fully intelligent. What we do argue is that having machines join humans in social construction processes holds implications for fundamental theoretical constructs we use both to build and explain societies. In particular, APs will raise new questions about influence, legitimacy, and agency.

\subsection*{Influence}

The prospect of having APs in social construction raises new questions about influence, a core topic of social science\cite{Cialdini}.  The idea of machines shaping social realities might seem far-fetched, but this is changing with the rise of social media influencers\cite{freberg2011social} and their digital doppelgangers, virtual influencers\cite{Arsenyan21VI,ContiVI22}. As virtual influencers spread and become more sophisticated, their growing efficacy will invite questions about the trustworthiness and ethics of their influence\cite{RobinsonVIs}.

But what capabilities will APs need to be more than mere marionettes? Participating in social construction processes need not require the stronger forms of power human actors use to dictate decisions and control agendas that determine what decisions can be made. Rather, engagement in social construction processes requires the subtler sort of influence that Lukes\cite{Lukes} calls the third form of power---the ability to shape what people believe is real, true, or right. As the ontological status of a potentially new social reality is being weighed and debated by competing voices, exactly who will hold sway over what people come to see as real is not known ex ante. Actors with traditional power bases are often outmaneuvered by innovators from the fringes of society. Even so, social theorists since Marx have linked this subtle form of power to social structures; Marx argued that the organization of industries reflects not just goods such as ‘‘cloth, linen, silks’’, but also the ‘‘ideas” and “categories” that establish this social order\cite{Marx}. Similarly, sociologist Charles Tilly observed that “concepts are tools” for enacting and naturalizing social fault lines capable of supporting persistently inequitable resource distributions\cite{Tilly}. 

Having APs shape the concepts and categories people use to make sense of and order the world will raise questions about how to assign responsibility, accountability, and liability for both good and bad outcomes of the socially constructed concepts that result. As it is with human influencers people, scholars and leaders of societies will grapple with how to separate scientifically sound solutions to fundamental problems from sensational social media mavens who misinform and mislead. This will be especially important since, like all technologies, APs could be developed to serve both these aims.

\subsection*{Legitimacy}

The prospect of APs also holds theoretical and practical implications for legitimacy, another core topic of social science. Legitimacy is an important construct because it exerts pressures to conform to beliefs, practices, and structures that become litmus for inclusion or exclusion in various segments of society. Also, legitimacy is a factor in the emergence and diffusion of concepts and categories that become fixtures in social ontologies\cite{TolbertandZucker}. The effects of legitimacy arise from two levels of judgment about social appropriateness: individuals' own beliefs and their estimates of what relevant collectives believe\cite{BitektineandHaack}. Crucially, legitimacy perceptions vary by audience, and in a world where digital technologies for learning offer a growing array of views, the legitimacy of even widespread social phenomena can vary widely from one audience, or subpopulation, to another.   

Legitimacy underlies all kinds of social evaluation, but it is especially important to reputations and shifts in criteria of reputations and their relative importance. As environmental, social, and governance (ESG) criteria have become important, for example, the former importance of maximizing shareholder value has declined. Organizational reputations have changed, and numerous new measures and practices are adopted. As AI systems evolve from mere compilation and analysis of organizational performance to also joining with human analysts in the identification and explanation of new patterns, the extent to which their growing participation in both social evaluation and standard setting becomes influential will raise questions about legitimacy that will require new ways of thinking about both legitimacy and reputation.

At the risk of oversimplifying, legitimacy is both a carrot and a stick: there are benefits to fitting what is legitimate and costs to flaunting it\cite{DiMaggioandPowell}. As virtual influencers spread, their potential influence and legitimacy will both reflect and constrain the impact they can have in social construction processes. They will also play a role in defining what is legitimate in society and this latter role has very significant societal implications.

\subsection*{Agency}

APs will also affect how both scholars and societies understand and assign agency---that is, the ability to think and act in ways that depart from commonly accepted beliefs, structures, norms, rules, or routines. Among academic definitions of agency, Sewell's\cite{Sewell} conceptualization of agency is particularly relevant to the participation game. For Sewell, agency is a "capacity to transpose and extend schemas to new contexts" that is "inherent in the knowledge of cultural schemas that characterizes all minimally competent members of society."\cite{Sewell} In folk wisdom, the individual is the locus of agency, but creating new social realities requires a collective response to new ways of seeing and doing things. Like Sewell, we see agency as inextricably related to the social structures that mediate collective agreements about what is real.

In the participation game, the prospect of agentic APs raises fundamental questions about accountability, liability, and freedoms. Even if the agency of APs is constrained by social structures in which they are embedded, as is the case for humans, conceptualizing AIs as having agency affects whether and how they will be held accountable for any influence they have. Also, the degree of agency APs have will affect whether and how they can receive credit for influence they have. In concept, this is akin to the handling of accountability for children. With young children, accountability for the deeds of children---both blame and credit---generally goes to parents or legal guardians. As children mature into independent young adults, however, accountability shifts away from parents and guardians and to the former children. 

Questions about the potential agency of APs also hold implications for the social structures these systems will be embedded in with their users and collaborators. As it is with humans, the agency of APs will reflect the constraints and degrees of freedom they have in selecting what ends to pursue. In more technical language, agency is about being able to change objective functions. Among humans, we see this as some people choose to play the game of life as given to them while others choose to play with the game in the hopes of creating worlds more to their liking. When it comes to these kinds of choices, some people have considerable freedom while others labour under burdensome constraints. The art and science of building AIs to collaborate with humans in social construction processes will raise questions about what kinds of inequalities and protections to engineer into human-AI collaborations that will shape social realities.    

\section*{Discussion: Implications for Society and AI Governance}

The participation game points the way to a new frontier for AI in which AI systems could evolve to participate in social construction processes. We argue that such an evolution will prompt re-thinking basic constructs of social theory; also, it will push societies to develop AI governance institutions AI that ensure safe deployment and use of new kinds of tools—or perhaps we should say “entities”—capable of shaping how people think. 

It is too early in the evolution of AI to say what system of governance will work best, but we can anticipate the problem and prepare by examining available models. Table 1 is a high-level summary and comparison of the rights, accountabilities, and freedoms associated with six different approaches to governing entities, both human and non-human, of varying abilities. Each model is an oversight relationship that links an overseer and the overseen and structures their respective rights and accountabilities. The rights column summarizes the rights of the overseen in each row; the accountabilities column characterizes the balance of accountability between the overseen and overseers in each row. For example, pets have limited rights and very few accountabilities, and people who keep pets have relatively light accountabilities for their care. In contrast, children have special rights that include some protections their overseers do not have. Parents face considerable accountabilities for meeting a standard of care for their children, bu children's accountabilities under the law are limited until they are of age. The governance models are listed in order of increasing rights and accountabilities for the overseen. Our main observation from Table \ref{tab:GovernanceModels} is that even with oversight of pets, pets’ overseers face non-trivial accountabilities for pet care. In contrast, the accountabilities of AI system developers are currently uncertain, and that will need to change as systems develop. 

\begin{table}[ht]
\centering
\begin{tabular}{|l|c|c|c|}
\hline
\emph{\textbf{Extant Governance Models} (o) overseen and (O) overseer} & \emph{Rights} & \emph{Accountabilities(o)} & \emph{Accountabilities(O)} \\
\hline
(o) Pets and their (O) owners & No & Low & Few \\
\hline
(o) Wards and their (O) conservators & Limited & Few & Modest \\
\hline
(o) Fictitious persons and their (O) limited-liability managers & Extensive & Considerable & Limited \\
\hline
(o) Children and their (O) parents & Special & Few & Considerable \\
\hline
(o) Talent and their (O) agents & Full Human & Diluted & Special \\
\hline
(o) Trainees and their (O) supervising professionals & Code & Profession & Membership \\
\hline
(o) Members of society and their (O) governing authorities & Full Human & Full Human & Sovereign \\
\hline
\end{tabular}
\caption{\label{tab:GovernanceModels} Extant Governance Models Potentially Applicable to Artificial Participants in Social Construction Processes}
\end{table}

Finally, the modern digital commons heralded in its early days as a meeting ground has evolved to  be a battle ground as well. To the extent AI systems have contributed to this evolution, we suggest the participation offers a hopeful post-Turing frontier for further AI evolution. There has been great value in imitation, but when imitation becomes deception, trust in science and technology is put at risk, and both science and its service to society are imperiled. In our view, it is time to shift focus from machine imitation of humans---too often deceptive---to constructive machine participation in quintessentially human projects.

\bibliography{ParticipationGame.bib}

\section*{Acknowledgements}

We thank Ron Chrisley and participants at the 2022 Oxford Corporate Reputation Symposium for their critiques and encouragement of this work.

\end{document}